\documentclass[doublecolumn]{aa} 
\usepackage{graphicx}
\usepackage{txfonts}

\usepackage[]{natbib}

\newcommand{\f}{4U\,1626--67 \,}

\newcommand{\swiftbat}{{\it Swift}/BAT }

\newcommand{\be}{\begin{equation}}
\newcommand{\ee}{\end{equation}}
\newcommand{\bdm}{\begin{displaymath}}
\newcommand{\edm}{\end{displaymath}}

\begin{document}
%

 \title{4U\,1626--67 as seen by \textit{Suzaku} before and after the 2008 torque reversal}

   \author{A. Camero--Arranz
          \inst{1}
          \and
          K. Pottschmidt\inst{2,3}
          \and
           M.H. Finger\inst{1,4}
           \and
          N.R.\,Ikhsanov\inst{5} 
          \and
          C.A. Wilson--Hodge\inst{4}
          \and
           D. M. Marcu\inst{2,3}
          }

   \institute{Universities Space Research Association (USRA), 6767 Old Madison Pike, Suite 450, Huntsville, AL 35806, USA, \email{camero@ice.csic.es}
             \thanks{Current address: Institut de Ci\`{e}ncies de l'Espai, ICE, (CSIC--IEEC), Campus UAB, 08193 Barcelona, Spain}
            \and
            NASA Goddard Space Flight Center, Astrophysics Science Division, Code 661, Greenbelt, MD 20771, USA
            \and 
            CRESST \& University of Maryland Baltimore County, 1000 Hilltop Circle, Baltimore, MD 21250, USA 
            \and            
            Space Science Office, ZP12, NASA/Marshall Space Flight Center, Huntsville, Alabama 35812, USA 
            \and            
            Pulkovo Observatory, 196140 St.\,Petersburg, Russia}

   \date{Received ; accepted }

 
  \abstract
   {}
   {The accretion--powered pulsar 4U\,1626-67 experienced a new torque reversal at the beginning of 2008, after about  18 years of steadily spinning down.  The main goal of the present work is to study this recent  torque  reversal that occurred in 2008 February.}
   {We present a spectral analysis of this source  using two pointed observations  performed by \textit{Suzaku} in 2006 March and  in 2010 September.  }
   {We confirm with \textit{Suzaku} the presence of a strong  emission-line complex centered on 1\,keV, with  the strongest line being the hydrogen-like Ne Ly$\alpha$ at  1.025(3)\,keV. We were able to resolve this complex with up to seven emission lines. A dramatic increase of the intensity of the Ne~Ly$\alpha$  line after the 2008 torque reversal occurred, with the equivalent width of this line reaching almost the same value measured by ASCA in 1993.  We also report on the detection of a cyclotron line feature centered at $\sim$37\,keV.  In spite of the fact that an increase of the X--ray luminosity (0.5--100\,keV) of a factor of $\sim$2.8  occurred  between these two observations, no significant change in the energy of the cyclotron line feature was observed.  However, the intensity of the $\sim$1\,keV line complex  increased by an overall  factor of $\sim$8.}
   {Our results favor a scenario in which the neutron star in 4U\,1626--67 accretes material from a geometrically thin disk during both the spin-up and spin-down phases.}


   \maketitle
%

\section{Introduction}

 \f  is a  Roche-lobe filling binary system, where matter from the companion flows from the L1 point into an accretion disk \citep{reynolds97}.  This low mass X--ray binary (LMXB)   consists of a 7.66 s X--ray pulsar accreting from an extremely low mass companion \citep[(0.04 M$\odot$ for {\it i} = 18$^o$, ][]{Levine88}.  Although orbital motion has never been detected  in the  X--ray data, pulsed optical emission reprocessed on the surface of the secondary revealed  the 42 min orbital period \citep{Middledich81}, confirmed by \cite{Chakrabarty98}.  Most likely the binary system  contains a hydrogen--depleted secondary to reach such a short orbital period \citep{Paczynski&Sienkiewicz81}. The faint optical counterpart (KZ TrA, {\it V}$\sim$17.5) has a strong UV excess and high optical pulse fraction \citep{McClintock77,McClintock80}.

 The sign of the torque in \f  has reversed  in only two occasions since its discovery by {\it Uhuru} \citep{Giacconi72}.  Monitoring of the source by the Burst and Transient Source Experiment (BATSE) on board the Compton Gamma Ray Observatory ({\it CGRO}) starting in  April 1991,  found the pulsar spinning down, implying a changed sign in the accretion torque \citep{wilson93,bildsten97a}. It was estimated that the reversal occurred in mid-1990.  More recently, our  daily monitoring of \f  with {\it Fermi}/GBM starting in 2008 August  found the pulsar again spinning--up rather than spinning--down. \swiftbat observations allowed us to cover  the evolution of this second torque reversal. The transition took place at around MJD 54500  (2008 Feb 04) and lasted  approximately  150 days \citep{camero10}.

A  48 mHz quasi-periodic oscillation (QPO) has been detected in the X--ray emission \citep{Shinoda90,KommersChakrabarty&Lewin98}. \cite{orlandini98}  inferred a neutron star magnetic field in the range of (2.4--6.3)$\times$10$^{12}$ G.  To compute this magnetic field range a source distance of 5--13 kpc was assumed \citep[and references therein]{Chakrabarty98}.  A $\sim$37 keV absorption cyclotron feature was found in the 0.1--200 keV  {\it BeppoSAX} spectrum \citep{orlandini98}.  A more detailed study of this feature  with \textit{Suzaku} can be found in \citet{iwakiri12}. The X--ray  broad-band continuum was fitted with  low--energy absorption, a blackbody, a power law and a high energy cutoff.

\citet{angelini95} using the ASCA observatory discovered for the first time a strong emission-line complex centered on 1\,keV in the X--ray spectrum of 4U\,1626-67. The strongest line was identified as hydrogen-like Ne  Ly$\alpha$ at 1.008\,keV. The strength of the neon emission compared to the expected iron L complex implied a large neon overabundance.

We present a spectral  analysis of \f with \textit{Suzaku}, using two observations from 2006  March and 2010 September. This  allows us to better characterize the recent  torque  reversal that occurred in 2008 February.


\begin{table*}
\begin{center}
\caption{Best Broad Band Fit  Spectral Parameters$^{a}$} 
\begin{tabular}{llllllllll}
  \hline\noalign{\smallskip}
 \hline\noalign{\smallskip}

  Observation            & $N_{\rm H}^b$  &  $\alpha^c$    & E$_{cut}^{d}$  &   E$_{Fold}^{d}$  &    E$_{CRF}^{d^*}$         &T$_{BBody}^{d}$           & Flux$^f$   & constant      &   $\chi_r^2$(DOF)  \\
                                 &                          &   norm             &                       &                          &     $\sigma_{CRF}^{d}$   & norm$_{BB}^e$               &                  &  factor$^{g}$ &  \\
  \hline \noalign{\smallskip} \hline \noalign{\smallskip}

    400015010            & 1.46(2)        & 0.846(9)        &      18.9(5)     & 13(1)              &  36.9$ {+0.9\atop -0.8}$ & 0.229${+0.01\atop -0.009}$ & 0.55(9)    &  0.85(2)    &1.38(205)  \\
   (Mar 2006)              &                   & 0.010(1)         &                        &                    & 3.6(6)                                 & 127$\pm$13                               &             &  0.8(fixed)     &            \\

  \hline\noalign{\smallskip}

   405044010             &  0.6(2)       &  0.981(3)        &     21.6(7)    &  15(2)             &  37.8(8)                              & 0.511(8)                                &  1.58(4)     &  1.21(4)    & 1.20(206)\\
    (Sep 2010)               &                &  0.026(1)       &                       &                       &  4.3(6)                      & 412${+146 \atop -108}$  &                 &  0.8(5)    &        \\
  \hline 

   \end{tabular}
 \end{center}
\begin{scriptsize}

\hspace*{2cm}    $^{a}$(PHABS (8~GAUSSIANS+GAUSSIAN+ BBODYRAD) +POW)*HIGHECUT *GABS)*CONST\\
\hspace*{2cm}     $^b  \times 10^{21}$\,cm$^{-2}$\\
\hspace*{2cm}     $^c$ Photon Index with normalization units in photons keV$^{-1}$cm$^{-2}$s$^{-1}$ at 1 keV.\\
\hspace*{2cm}     $^{d}$  keV\\
\hspace*{2cm}     $^{d^*}$ The values for the cyclotron line depths are 14$\pm$4 and  18$\pm$4 for the Mar 2006 and  Sep 2010 observations, respectively.\\
\hspace*{2cm}     $^{e}$ with norm$_{BB}$=R$_{BB_{km}}^2$/d$_{10}^2$, where d$_{10}$ is the distance  in units of 10 kpc, varying from 0.5-1.3 (assuming the 5 --13 kpc distance range reported for this source).\\
\hspace*{2cm}     $^{f}$ $\times 10^{-9}$\,erg\,cm$^{-2}$\,s$^{-1}$    (0.5--100 keV), based on the XIS\,03 flux calibration; XIS\,03  is a combined spectrum  from XIS\,0 and XIS\,3.\\
 \hspace*{2cm}    $^{g}$ Intercalibration factors (PIN/XIS\,03 and GSO/XIS\,03). The GSO/XIS\,03 factor was fixed to   the value obtained for the second observation since it could not be well \\  \hspace*{2.1cm}  constrained.\\

\label{specfits}
\end{scriptsize}
\end{table*}



\section{Observations and Analysis}

The instruments on \textit{Suzaku} \citep{mitsuda07} are the X-ray Imaging Spectrometer \citep[XIS;][]{koyama07} CCD detector covering the  0.3--10 keV band, and the Hard X-ray Detector \citep[HXD;][]{takahashi07} comprised of  the PIN diode detector (PIN) covering the 10--70 keV band  and the gadolinium silicate crystal detector (GSO) covering  the 60--600 keV band. The XIS has four separate detectors, XIS\,0--3, with XIS\,1 being a backside illuminated CCD and the others frontside.  XIS\,2 was lost due to a micrometeor hit in late 2006.

The first \textit{Suzaku} \f  observation occurred in 2006 March (MJD 53797) and thus before the torque reversal in 2008.
During this observation the  spacecraft aimpoint was optimized for the XIS detectors, and \textit{Suzaku} was run in a  1/8 window mode with 1\,s integration time, an effective area of $128\times1024$ pixels and an effective exposure time of $\sim$72.2\,ks. The second observation in 2010 September (MJD 55444) was also centered on the XIS aimpoint but run in a 1/4 window mode, with 2\,s integration time, an effective area of $256\times1024$ pixels and an effective exposure time of $\sim$14.01\,ks. For this observation only 3 XIS detectors were available,  as we discussed in the previous paragraph. The \textit{Suzaku} data were reduced with tools from the HEASOFT v6.11 package \citep{arnaud96} and CALDB release 20110804.  Following the \textit{Suzaku} ABC guide\footnote{http://heasarc.nasa.gov/docs/suzaku/analysis/abc}, the first step of our analysis was to reprocess the data by running the FTOOL \textit{aepipeline} for the XIS, the PIN and the GSO.  The pipeline performs calibration as well as data screening.


Due to thermal distorsion of the \textsl{Suzaku} spacecraft the optical bench shows wobbling. The source's PSF image is not centered on a fixed position on the XIS CCDs but appears blurred. As part of the standard data reduction this effect is partially corrected by adjusting the spacecraft attitude file. The attitude file can be further improved by applying the \texttt{FTOOL} \texttt{aeattcor2}: based on the algorithm described by Nowak et al. (2011) the mean event positions as a function of time are calculated and used to further adjust the attitude. While we obtain sharp images for our first \textsl{Suzaku} observation, the second one suffers from additional attitude variability that can be present in \textsl{Suzaku} observations since a change in the attitude control system on 2009 Dec 18. In this case the \texttt{FTOOL} cannot fully correct the characteristic "double images". According to the documentation\footnote{ftp://legacy.gsfc.nasa.gov/suzaku/doc/general/suzakumemo-2010-05.pdf}, however, and since conservative pileup exclusion regions were applied (see below), the XIS spectra are not significantly affected.

We used the \texttt{FTOOL} \texttt{pileest} to estimate the amount of pileup in the XIS images and disregarded circular regions with $>8$\%
pileup fraction during the spectral extraction. This is a careful approach since due to the asymmetric shape of the PSF only few pixels outside of the exclusion region reach pileup fractions above $2-4$\%. For the first observation the pileup fraction remained below this limit in the entire source extraction region (a circle of $\sim70^{''}$ radius centered on the source image). In the second observation pileup fractions of up to 15$\%$  were reached in the center of the source image so the extraction region consisted of a circle of  $\sim130^{''}$ radius minus a central circular exclusion region with a radius of  $\sim35^{''}$.

\begin{figure*}[t]
\hspace{-0.5cm}
\includegraphics[width=9.3cm,height=11.5cm]{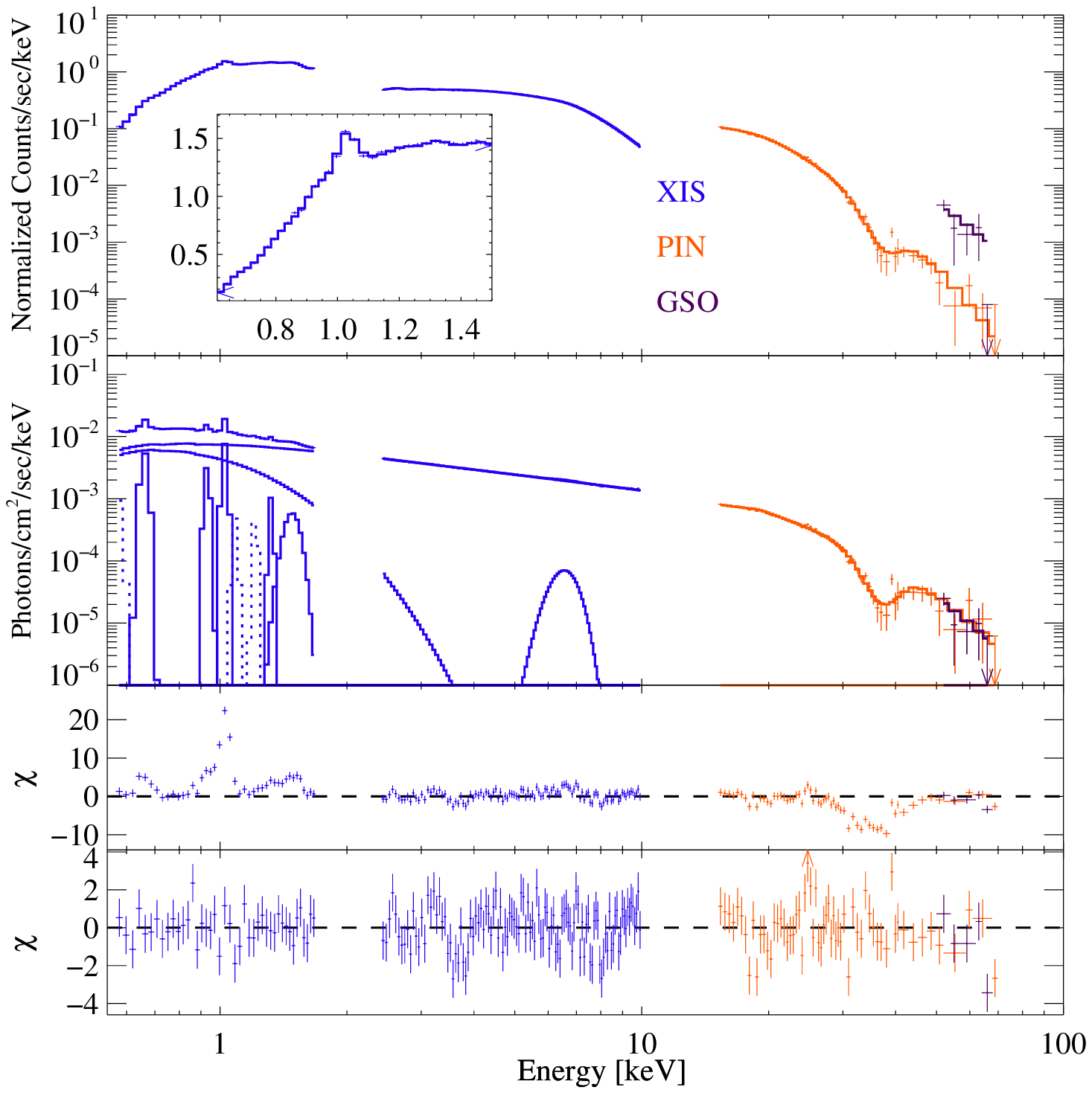}
\includegraphics[width=9.4cm,height=11.5cm]{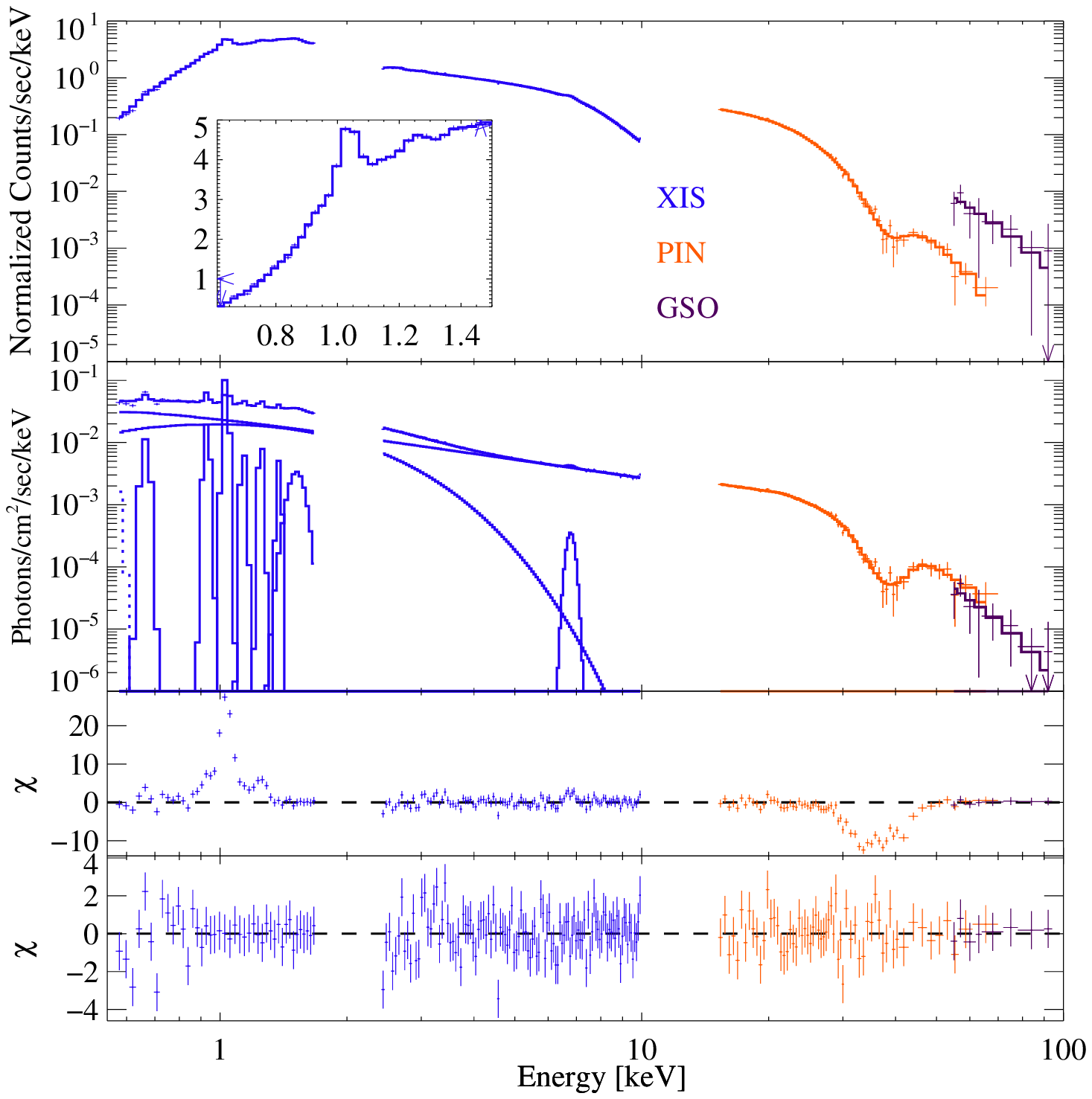}
 \caption{The main panels show the \textit{Suzaku} broad-band spectrum before (left panel) and after (right panel) the 2008 torque reversal.  Blue points denote XIS data (0.5--10\,keV), PIN data points are in red (14--70\,keV) and GSO in purple (50--70\,keV and 50--100\,keV, respectively). The residuals after fitting only the continuum model are shown in the middle panel, with no line emission  and cyclotron line feature included. The bottom panel shows the residuals after fitting the line complex at $\sim$1\,keV, the $\sim$6.5\,keV iron line and the $\sim$37\,keV cyclotron line (see Table~\ref{specfits} for a list of fitted parameters). A zoom of the region of the line complex at $\sim$1\,keV  is also shown for clarity.\label{spec}}
\end{figure*}


\begin{figure*}[t]
\hspace{-0.7cm}
\includegraphics[width=9.2cm,height=8.5cm]{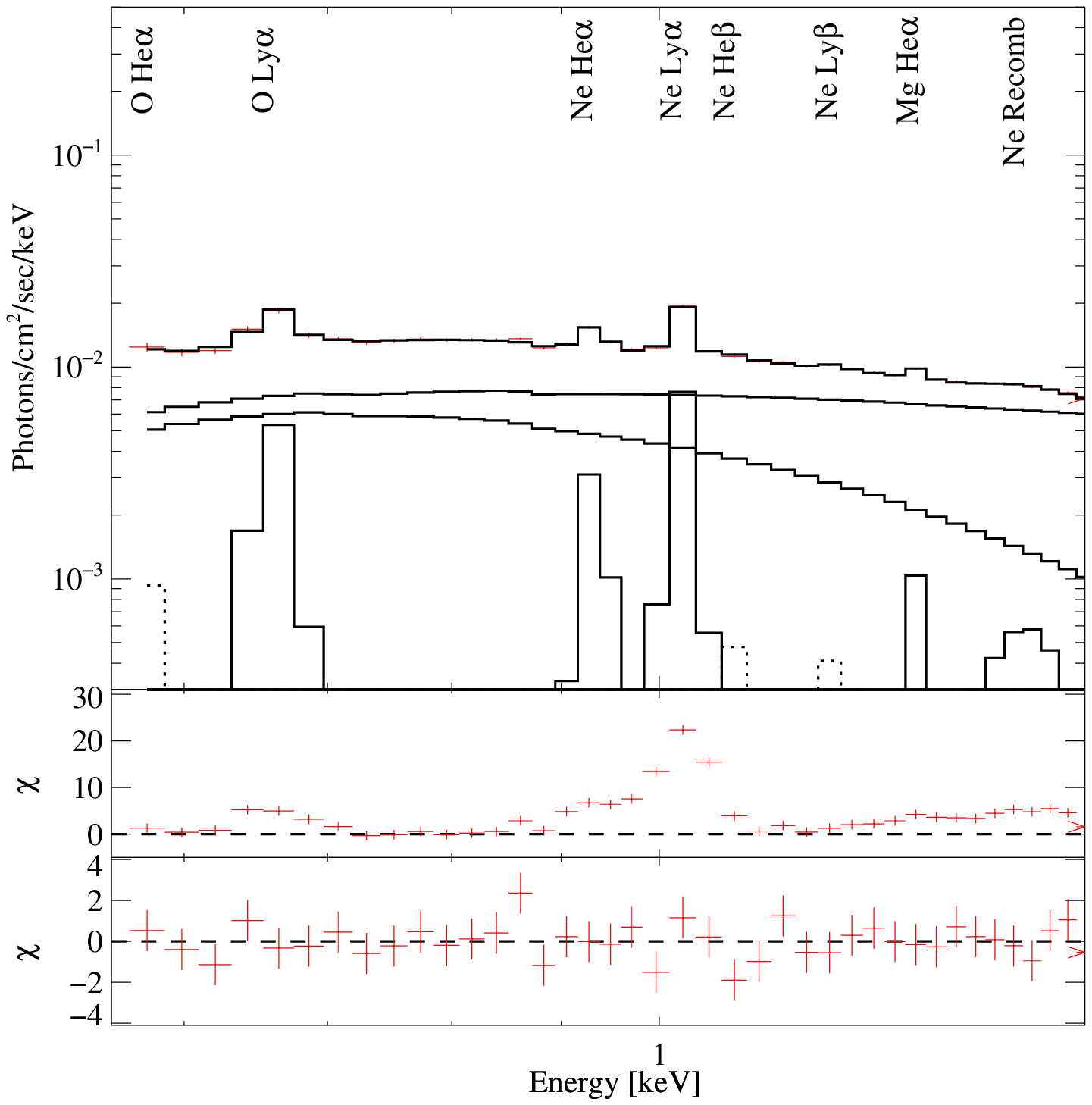}
\includegraphics[width=9.4cm,height=8.5cm]{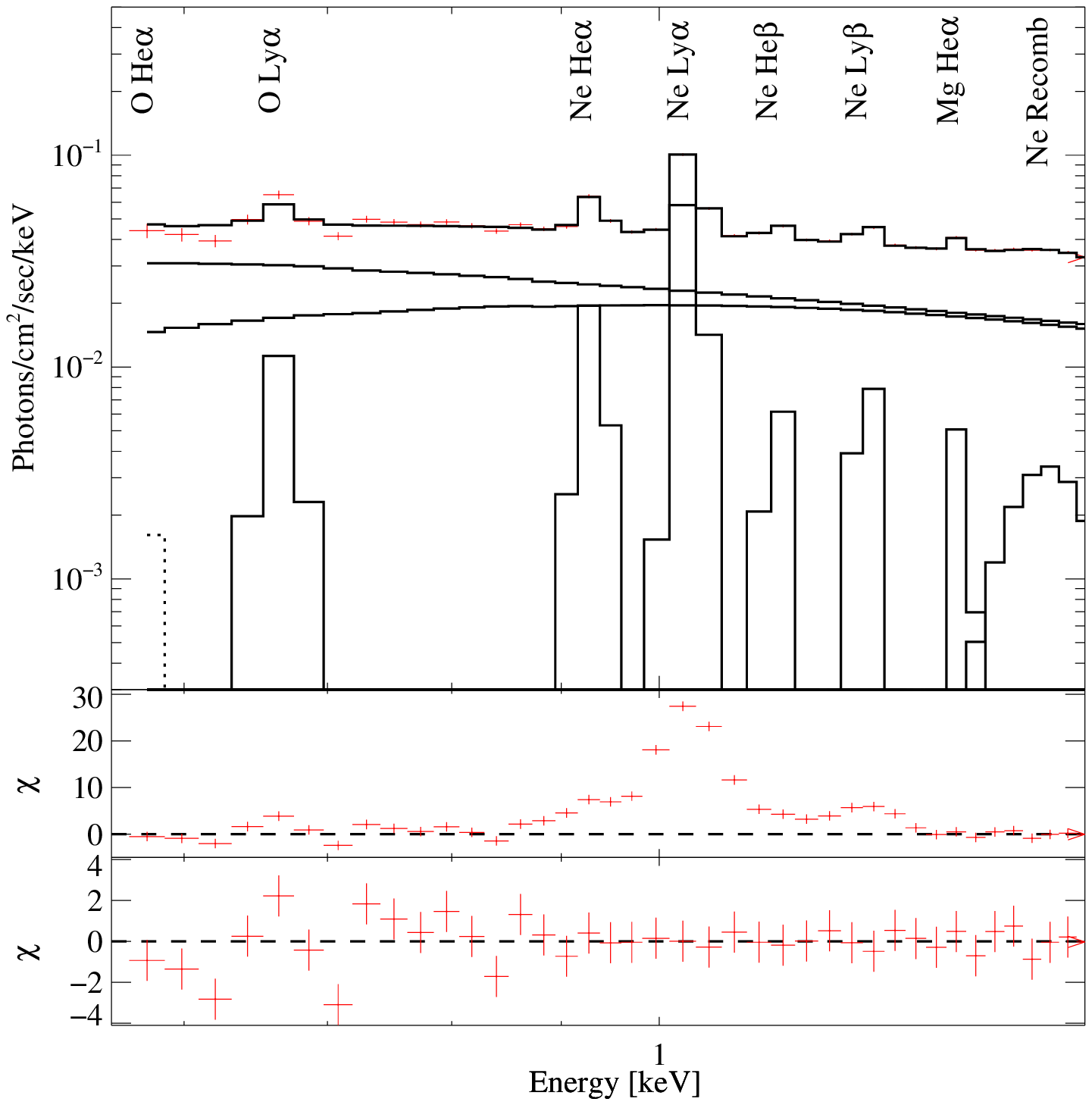}
 \caption{The top panel shows the 8 Gaussian lines from the $\sim$1\,keV complex detected by  \textit{Suzaku},  before (left panel) and after (right panel) the 2008 torque reversal.  Here the dotted lines in the model denote the components that were fixed.  The residuals after fitting only the continuum model are shown in the middle panel, with no line emission  included. The bottom panel shows the residuals after fitting the line complex with those 8 lines (see Table~\ref{linesfit} for more information about these lines). 
 \label{lines}}
\end{figure*}

The primary tool for extracting data products (spectra, lightcurves) from XIS data is \textit{xselect}, which is part of the general HEAsoft distribution.  Events in the XIS detectors were in the 3$\times$3 and 5$\times$5 editing modes. With \textit{xselect} we applied to the cleaned events filters with the good times intervals avoiding telemetry saturation. Source and background regions were created from the image. The filtered events were then used to create source and background spectra for the desired extraction regions. Products from the two editing modes were added. The XIS spectra were grouped close to the energy resolution of the detectors \citep[see][]{nowak11}. We  created response matrices and effective area files  with the \textit{xisrmfgen} and \textit{xissimarfgen} tools, respectively.  The calibration of the XIS is uncertain around the Si\, K edge, thus  events in the 1.8--2.0 keV range were excluded in the fitting. For more information about this issue we refer the reader to the "XIS Calibration around the Si K edge" web page\footnote{http://heasarc.gsfc.nasa.gov/docs/suzaku/analysis/sical.html} \citep[see also][for different treatments on this matter]{iwakiri12, nowak11, suchy11}. 

For the PIN, we extracted spectra from the cleaned event files using \textit{hxdpinxbpi}. This tool produces the dead time corrected PIN source spectrum as well as the PIN background spectrum. The non X--ray background events provided by the \textit{Suzaku} team  can be downloaded from  the HEASARC  archive\footnote{ftp://legacy.gsfc.nasa.gov/suzaku/data/background/pinnxb$\_$ver2.0$\_$tuned}.  The PIN spectra were grouped to have a signal-to-noise ratio $\gtrsim$ 10 in each energy bin, and we considered spectra between 14--70\,keV. We point out that due to the changes in instrumental settings the PIN response matrices, from the CALDB database, are different depending on the epoch of the observation.

The GSO spectra were created starting with the clean event files using the \textit{hxdgsoxbpi} tool.
The background events were downloaded from HEASARC\footnote{ftp://legacy.gsfc.nasa.gov/suzaku/data/background/gsonxb$\_$ver2.5}. Response files were  taken from the CALDB database and the GSO correction ARF file was provided by the HXD instrument team\footnote{http://www.astro.isas.ac.jp/suzaku/analysis/hxd/gsoarf/}. The distributed GSO Non X--ray Background is grouped, so GSO spectra have to be grouped accordingly, using the spectral binning given by the file \textit{gsobgd64bins.dat} available at HEASARC\footnote{ftp://legacy.gsfc.nasa.gov/docs/suzaku/analysis/gsobgd64bins.dat}.  The GSO spectrum  was restricted to the 50–-70\,keV range for the first observation, and  to the 50--100\,keV for the second observation.


\begin{table*}[]
\begin{center}
  \caption{Fit  Results on Line Emission$^a$} 
 \begin{tabular}{llllllllll}
  \hline\noalign{\smallskip}\hline\noalign{\smallskip}

                              &\hspace{-0.35cm} Expected    & \hspace{0.2cm} Line Energy&\hspace{-0.35cm}(keV)\hspace{0.3cm} &  \hspace{0.6cm}      $\Delta~\chi^2$&            & \hspace{0.4cm} Intensity$^b$ &  &   \hspace{0.6cm}  EW   &\hspace{-0.5cm}   (eV)   \\

Observation           &\hspace{-0.35cm}  (keV)      &  Mar 2006 &\hspace{-0.35cm}  Sep 2010  \hspace{0.3cm}  &Mar 2006 &\hspace{-0.25cm} Sep 2010   &  Mar 2006  & \hspace{-0.35cm} Sep 2010 \hspace{0.3cm}  &  Mar 2006 &\hspace{-0.35cm}  Sep 2010  \\

&&&&(dof=205)&\hspace{-0.35cm}  (dof=206)&&&&\\
\hline\noalign{\smallskip}\hline\noalign{\smallskip}

O~He$\alpha$          &\hspace{-0.35cm} 0.568       &   0.568(fixed)   &\hspace{-0.35cm}   0.568(fixed) & --  &\hspace{-0.35cm} --        & 0.09(fixed)          &\hspace{-0.35cm}  0.09(fixed) 	  &   3.2(2)    &\hspace{-0.35cm}  1.3(2)\\

O~Ly$\alpha$           &\hspace{-0.35cm} 0.653      &   0.661(8)       &\hspace{-0.35cm}   0.66(2)      & 59  &\hspace{-0.35cm} 5.4$^c$      & 3.0${+1\atop -0.9}$&\hspace{-0.35cm}     4(4)          &    13(4)   &\hspace{-0.35cm}   7(3) \\

Ne~He$\alpha^d$     &\hspace{-0.35cm} 0.914         &   0.930(8)       &\hspace{-0.35cm}   0.929(8)     & 105 &\hspace{-0.35cm} 119     &   1.3(3)      	              &\hspace{-0.35cm}     7(1)          	&    8(1)     &\hspace{-0.35cm}   14(1)      \\

Ne~Ly$\alpha$         &\hspace{-0.35cm} 1.021       &   1.025(3)       &\hspace{-0.35cm}   1.031(2)     & 878 &\hspace{-0.35cm} 1699     &    3.3(2)  		           &\hspace{-0.35cm}    24(1)         &   22.6(2)  &\hspace{-0.35cm}   50.9(8)  \\

Ne~He$\beta^d$      &\hspace{-0.35cm} 1.084         &   1.084(fixed)   &\hspace{-0.35cm}   1.13(3)      & --  &\hspace{-0.35cm} 39      &    0.2(fixed)              &\hspace{-0.35cm}     3(1)     		&   3.7(2)   &\hspace{-0.35cm}   6(2)     \\

Ne~Ly$\beta$          &\hspace{-0.35cm} 1.210       &   1.210(fixed)   &\hspace{-0.35cm}   1.25(1)      & 6   &\hspace{-0.35cm} 96       &    0.2(fixed)               & \hspace{-0.35cm}    2(1)     		&  1.8(2)    &\hspace{-0.35cm}   9(4)   \\

Mg~He$\alpha$       &\hspace{-0.35cm} 1.34          &   1.32(2)        &\hspace{-0.35cm}    1.38(2)     & 36  &\hspace{-0.35cm} 31     &   0.4(1)              &\hspace{-0.35cm} 1.8${+0.9\atop -1}$&  4(1)   &\hspace{-0.35cm}   5(1)   \\

Ne Recombo$^e$     &\hspace{-0.35cm} 1.362          &   1.48(2)        &\hspace{-0.35cm}    1.52(2)     & 137 &\hspace{-0.35cm} 154     &   0.9(2)                       &\hspace{-0.35cm}     5(1)     		 &  11.1(3)  &\hspace{-0.35cm}   15(5)  \\

Fe K-fluor. line$^f$ &\hspace{-0.35cm} 6.4          & 6.5(2)           &\hspace{-0.35cm}    6.78(7)     &  59  &\hspace{-0.35cm}  30   &   0.8$ {+0.4\atop -0.3}$&\hspace{-0.35cm}  1.3(5)       & 42.2(3)   &\hspace{-0.35cm}  33(1)\\
 \hline

 \end{tabular}  \label{linesfit}
\end{center}
\begin{scriptsize}
\hspace*{0.2cm}   $^a$  the widths of the low-energy lines were fixed to the ASCA values (10~eV for all but the 1.362~keV line with 57~eV)\\
\hspace*{0.2cm}   $^b$  photons 10$^{-4}$cm$^{-2}$s$^{-1}$\\
\hspace*{0.2cm}   $^c$  we note that,  for the 2010 observation, the $\Delta~\chi^2$ for this line is not well computed due to the spectral model not being able to adequately  fit this line (see Fig.~\ref{lines}, right panel).\\
\hspace*{0.2cm}   $^d$  the abundances of the lines at 0.914 and 1.084 keV can also be consistent with a blend of several \textit{L}-shell iron line emission (see Angelini et al. 1995).\\
 \hspace*{0.2cm}  $^{e}$ the Mg H$\alpha$ line emission occurs at 1.471 keV (here not resolved, and thus not detected, by \textit{Suzaku}), and therefore this Mg line may be contributing to this feature.\\
 \hspace*{0.2cm}   $^f$ for the iron line the widths were fitted (0.5(3) keV and  0.14(7) keV, respectively) \\
 \end{scriptsize}
\end{table*}


\section{Results}

The $\sim$0.5--100 keV broad-band spectrum obtained using XIS (a combination of detectors 0 and 3), PIN and GSO were fitted in XSPEC\,12.7.0.  We used a canonical model which  allowed us to compare our spectral study  with previous works \citep[among others] {angelini95,pravdo79,orlandini98,krauss07,jain10} and update the  long-term  X--ray flux history of 4U\,1626--67  relative to the flux measured by {\it HEAO 1} \citep{chakrabarty97, krauss07, camero10}.  

The continuum model fitted to both observations includes low-energy absorption, a blackbody component, and a power law with a high-energy cutoff at  $\sim$20 keV  (PHABS (BBODYRAD+POWLAW) HIGHECUT)*CONST. The constant factor (CONST) is used to account for intercalibration  between XIS, PIN, and GSO. The abundance was set to wilm (Wilms et al. 2000). Spectral fits for both observations are shown in Figure~\ref{spec}. A broad iron line near 6.4 keV (GAUSS) and a $\sim$37\,keV cyclotron feature (GABS) improved the fit for both observations \citep[see also][]{iwakiri12}. Table~\ref{specfits}  presents the fitted values for these broad-band parameters from the final model (which included the low-energy lines described below). 

Initial fitting results for the broad-band model for both observations indicated evidence for a strong line complex at $\sim$1\,keV, shown in Figure~\ref{spec}, middle panels. Following Angelini et al. (1995), we added seven narrow Gaussian lines (listed in Table~\ref{linesfit}) to our model. The line widths were fixed at the ASCA values (10 eV for all lines except the  1.362~keV  line width was fixed at 57~eV). Table~\ref{linesfit} lists the final fitted line parameters for both  observations. Only five of the seven lines detected with ASCA were detected in the 2006 \textit{Suzaku} observation, while six ASCA lines were detected in the 2010 \textit{Suzaku} observation. For the tree ASCA lines not well detected in 2006 and the one in 2010  the intensities were fixed,  since the errors could not be well constrained.

The broad feature at 1.362~keV  could alternatively be explained by a recombination edge. Fits to both observations using a model that replaced the 1.362~keV  Gaussian line with a recombination edge yielded similar $\delta \chi^2$  to our model containing the 1.362~keV line. The edge (and temperature) obtained for the 2006 and 2010 observations are  1.45$\pm$0.05~keV (42${+50\atop -40}$~eV)  and 1.47$\pm$0.03~keV (45${+34\atop -25}$~eV) (fitting with a Gaussian line gives 1.48(2) and 1.51(2)\,keV, see Table~\ref{linesfit}), respectively. A posteriori inspection of the residuals for both observations, suggested the presence of an extra Gaussian component at energy $\sim$1.34\,keV. Therefore, up to 8 Gaussian components were included in the fit (see Fig.~\ref{lines}; $\Delta~\chi^2$ of 283 for 205 dof, and 248 for 206 dof, for the first and second observations, respectively).  Final fit parameters for this line at $\sim$1.34\,keV, tentatively identified as the He-like Mg, are listed in Table~\ref{linesfit}.  The $\Delta~\chi^2$ obtained by setting the flux of a particular line to zero in the best-fit model  indicates the significance of each line. In addition, we  fit the broad-band spectrum using different combinations of the XIS detectors yielding similar results.  Other models accounting for emission  from a hot diffuse gas, with line emissions coming from several elements, were also fit (\textit{meka} and \textit{vmeka} in XSPEC notation).  We changed the  elemental abundances while performing the fits with the \textit{meka} and \textit{vmeka} models, however they did not  give reasonable fits.

A  luminosity increase occurred between the two observations, changing from L$_{(0.5-10\,keV)}\approx$2.4$\times$10$^{34}$d$_{kpc}^2$ to 6$\times$10$^{34}$d$_{kpc}^2$ erg s$^{-1}$.  However, the value of the center of energy of the cyclotron line  did not exhibit a significant change. This behavior resembles the  case of the Be/X-ray binary system A\,0535+26 \citep{caballero09}, but is contrary to sources like Her\,X--1 and V\,0332+53 \citep[and references therein]{pottschmidt11} or GX\,304--1  \citep{klochkov12}, which show  positive or negative L--E$_{\rm~ CRF}$ correlations \citep[see also][]{becker12}. On the other hand, the  column density, $N_{\rm H}$, dropped by a factor of $\sim$3 while the black body temperature increased by a factor of $\sim$2. The rest of the broadband parameters were consistent between the two observations. Absorbed fluxes in the 0.5--10\,keV, 2--10\,keV,  2--20\,keV and 2-50\,keV bands are 0.18(4), 0.159(2), 0.295(3), 0.51(3)~$\times$10$^{-9}$ erg cm$^{-2}$s$^{-1}$ and  0.49(3),  0.38(9),  0.64(1),  1.3(9)~$\times$10$^{-9}$  erg cm$^{-2}$s$^{-1}$,    for the 2006 and 2010 \textit{Suzaku} observations, respectively.

\section{Discussion and Conclusions}

\subsection{Spectral Continuum}

 Our results show that  two years after the torque reversal in 2008  the  luminosity has reached almost the same level as in 1977 (see Fig.~\ref{fluxhist}), the black body temperature has increased to $\sim 0.6$\,keV and the pulsar shows the same spin-up rate. The system's appearance in this state can be explained in terms of accretion onto the neutron star from a geometrically thin Keplerian disk provided the mass-transfer rate is limited to $\dot{M} \geq 10^{16}\,{\rm g\,s^{-1}}$ \citep{chakrabarty97}.

The pulsar's transition from spin-up to spin-down in 1990 was accompanied by significant changes of spectral parameters. The black body temperature decreased from $\sim 0.6$\,keV to $\sim 0.3$\,keV and the power law photon index changed from 
$\sim$1.5 \citep{vaughan97} to  $\sim$0.6 \citep{vaughan97,angelini95,owens97}. The transition period lasted about 150\,days and the source luminosity during this time was only a factor of~2 smaller than its highest level observed in 1977 (see Fig.~\ref{fluxhist}). This indicates that the mass-transfer rate during the torque reversal in 1990 was in excess of $5 \times 10^{15}\,{\rm g\,s^{-1}}$.

Our \textit{Suzaku} observations show that the pulsar's behavior during the torque reversal in 2008 (from spin-down to spin-up phase) was different. While the transition period also lasted about 150\,days and the black body temperature changed from $\sim 0.23$\,keV back to $\sim0.6$\,keV in 2010, the photon index remained almost unchanged. The value of the order of unity observed in 2006 and 2010 is the same as the photon index observed by several missions before the torque reversal during the spin-down phase, e.g. \textit{ASCA} \citep{angelini95}, \textit{BeppoSAX} \citep{orlandini98}, \textit{Chandra} and \textit{XMM} \citep{krauss07,schulz01}. The torque reversal in 2008 occurred at slightly lower luminosity level than in 1990 and was followed by a brightening of the X-ray source on a time scale of more than 2~years.

As we have shown in a previous paper \citep{camero10}, the torque reversals in \f are unlikely to be associated with alternation of the accretion flow geometry between the Keplerian geometrically thin disk (during the spin-up phase) and a spherical flow or a hot, geometrically thick disk (during the spin-down phase). The corresponding scenarios \citep[see, e.g.][]{yi99} encounter major difficulties explaining 150\,day transition period \citep[see also][]{WijersPringle99} and the low level of X-ray flux measured before and after the torque reversal in 2008. Furthermore, the mass-transfer rate during both of the observed torque reversals ($\geq 5 \times 10^{15}\,{\rm g\,s^{-1}}$) was above the critical mass-transfer rate at which the transition of the accretion disk to the hot, geometrically thick state can be expected \citep{Narayan-Yi-1995}. Finally, observations of the strong emission-line complex centered on 1\,keV and the iron K-fluorescence line at 6.4\,keV during both spin-up and spin-down phases indicate that the temperature of the material surrounding the neutron star does not exceed a few keV. Conversely, the temperature of the material in a thick disk is close to the adiabatic temperature, which for the parameters of 4U\,1626--67 is in excess of 10\,keV. Thus, our results favor a scenario in which the neutron star in 4U\,1626--67 accretes material from a geometrically thin disk during both the spin-up and spin-down phases.

The magnetospheric radius of a neutron star accreting material from a geometrically thin disk can be evaluated as $r_{\rm m} \simeq \kappa \left(\mu^2/\dot{M} (2GM_{\rm ns})^{1/2}\right)^{2/7}$, where $\mu = (1/2) B_* R_{\rm ns}^3$ is the dipole magnetic moment, $B_*$ the surface field and $R_{\rm ns}$ the radius of the neutron star. $\kappa$ is a parameter accounting for the geometry of the accretion flow, which in the case of disk accretion is limited to $\kappa \geq \sim 0.5$ \citep{ghoshlamb79, wang96, bozzo2008}. For the star to be in the accretor state its magnetospheric radius should not exceed the corotation radius, $r_{\rm cor} = \left(GM_{\rm ns} P_{\rm s}^2/4 \pi^2\right)^{1/3}$. Otherwise, the centrifugal barrier at the magnetospheric boundary would prevent the accretion flow from reaching the stellar surface \citep{Shvartsman-1970}. This condition in the case of \f is satisfied if the mass-transfer rate is $\dot{M} \geq \dot{M}_{\rm min}$, where
 \be\label{mdotmin}
 \dot{M}_{\rm min} \simeq 6 \times 10^{14}\,{\rm g\,s^{-1}}\ \kappa_{0.5}^{7/2} m^{-5/3} P_{7.7}^{-7/3} \mu_{30}^2
  \ee
Here $\kappa_{0.5} = \kappa/0.5$ is normalized to its minimum value, $m$ is the mass and $P_{7.7}$ the spin period of the neutron star in units of $1.4\,M_{\sun}$ and 7.7\,s. The dipole magnetic moment of the neutron star, $\mu_{30}$,  is normalized to $10^{30}\,{\rm G\,cm^3}$. According to our measurements of the energy of the cyclotron line feature with \textit{Suzaku} (36.8${+0.9\atop -0.8}$\,keV) the surface field of the neutron star in \f is  $\sim 3 \times 10^{12}$\,G. This is consistent with
previous estimate of the field strength made by \citet{orlandini98} from observations of \f with \textit{BeppoSAX}. Putting this value to Eq.~(\ref{mdotmin}) and taking into account that \f is a persistent pulsar one finds that the mass-transfer rate has never decreased below $10^{15}\,{\rm g\,s^{-1}}$ in the time period between 1977 and 2010. The rate of mass-transfer, therefore, may vary in the system up to an order of magnitude. It is remarkable that the observed X-ray flux (and, correspondingly, the mass accretion rate onto the surface of the neutron star) varies with similar amplitude (see Fig.~\ref{fluxhist}). This indicates that either the value of $\kappa$ is indeed close to its lower limit, or the flux variations are governed by a different reason. In particular, one can also envisage a situation in which the magnetospheric radius of the neutron star during the low luminosity phase is close to the corotation radius (this implies $\kappa$ to be substantially larger than its minimum possible value) and the centrifugal barrier at the magnetospheric boundary reduces the mass transfer rate into the stellar magnetosphere (for discussion see Perna et al. 2006). The mass-transfer rate in the disk within this scenario varies insignificantly as the pulsar switches between the spin-up and spin-down phases.

While the black body temperature during 2006-2010 increased from 0.23 keV to 0.51 keV, the source luminosity increased only by a factor of $\sim$2. This implies that during the brightening phase the effective {area of the black-body source has decreased by almost an order of magnitude. A high luminosity and a relatively small size of this source suggest that it is located inside the magnetosphere in the vicinity of the neutron star. The accretion process in this region is fully controlled by the stellar magnetic field and plasma flows within the accretion channel. In this light, variations of the effective area of the black-body source can be associated with a restructuring of the accretion channel (i.e. change of both its cross-section and the height of the shock at the base of the accretion column). Studies \citep[see e.g. ][]{burnard83,basko76,kulkarni_romanova08} suggest that the parameters of the channel depend on the geometry and physical conditions in the accretion flow at the magnetospheric boundary, as well as on the mode by which the material enters the magnetic field of the neutron star. Our observations give no evidence of any significant change in the accretion flow geometry during the brightening phase (see above). It, therefore, appears more plausible to associate the restructuring of the accretion channel with change of the mode by which the accreting material enters the magnetosphere.

It is widely adopted that the accreting material can enter the magnetosphere due to the Rayleigh-Taylor and Kelvin-Helmholtz instabilities or/and the magnetic reconnection. The Rayleigh-Taylor instability can be at work if the angular velocity of the accreting material at the boundary does not significantly differ from the angular velocity of the neutron star itself. The magnetic field configuration in the magnetopause in this case is close to poloidal. If this condition is not satisfied the Rayleigh-Taylor instability can be suppressed by the magnetic field shear in the magnetopause \citep{burnard83,ikhsanov96}. This indicates that the relative velocity between the magnetic field of the neutron star and the accreting material at the boundary can be one of the key parameters determining the structure of the accretion channel. Even small fluctuations in the mass transfer rate may lead to significant variations of this parameter if the magnetospheric radius is close to the corotation radius \citep[see e.g.][]{perna06}. Another possibility is an amplification of the magnetic field at the inner radius of the disk governed by the dynamo action \citep[see e.g. ][]{Rappaport04}. The disk in this case may switch into a sub-Keplerian state and decrease the angular velocity of the material at its inner radius. However, the modeling of the evolution of the channel, and its influence on the properties of the black body source within these scenarios, is very complicated and beyond the scope of the present paper.

\begin{figure}
\begin{center}
\includegraphics[width=8.75cm,height=7cm]{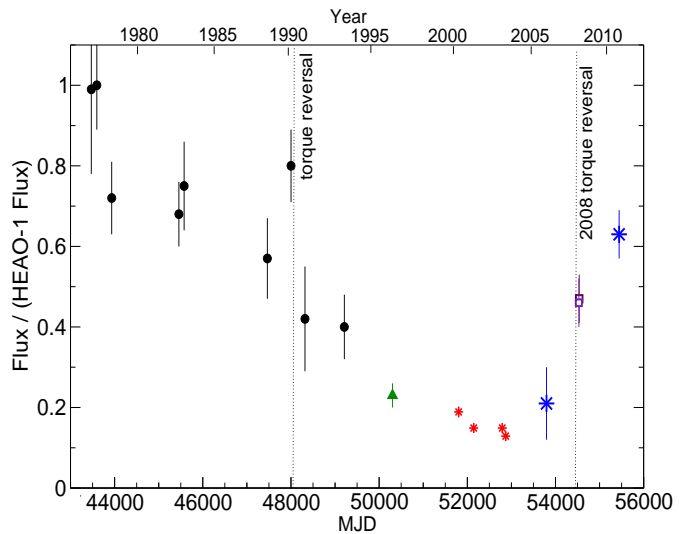}
 \caption{The X--ray flux history of 4U 1626--67  relative to the flux measured by {\it HEAO 1},
in the same energy band, from previous works (Chakrabarty et al. (1997): circles; Orlandini et al.  (1998): triangle; Krauss et al. (2007): stars; Camero-Arranz et al. (2011) unfilled squares) and the two \textit{Suzaku} observations (blue diamonds) in the 0.7--60 keV band. \label{fluxhist}}
\end{center}
\end{figure}


\subsection{Emission Lines}

The observed change of the equivalent width (EW) of the  Ne~Ly$\beta$ line at 1.021 keV during the brightening phase of the pulsar from 22.6 to 50.9\,eV, as well as the increase of its intensity  a factor of $\sim$8  reflect a transition of the accretion picture to a new phase. This may indicate either  changes of plasma parameters at the inner radius of the accretion disk \citep{kii86} or changes in the wind caused by interaction between the Keplerian disk and the stellar magnetic  field at the corotation radius \citep{lovelace95}. As shown by \citet{perna06}, the mass outflow rate can be higher during the spin-down  phase and lead to a periodic circulation of material between the inner and outer radii of the accretion disk. This feedback results in a redistribution of mass in the accretion disk and can lead to variations of the mass-transfer rate in the system. If this is the cause of the observed luminosity increase after the torque reversal in 2008, the time scale of radial motion of the material in the disk is $\tau_{\rm r} \sim 2$~years. This situation can be realized if the $\alpha$ parameter at the outer radius of the disk is  relatively small,
 \be
\alpha \sim 0.02\ m^{1/2} \left(\frac{c_{\rm s}}{10^6\,{\rm cm\,s^{-1}}}\right)^{-2} \left(\frac{\tau_{\rm r}}{2\,yr}\right)^{-1} \left(\frac{R}{10^{10}\,{\rm cm}}\right)^{1/2},
\ee
and, correspondingly, the density of material in the disk significantly exceeds its average value for X-ray pulsars \citep[for discussion see e.g.][]{king12}.  Here $c_{\rm s}$ is the sound speed in the disk at the distance $R$. The slow turbulization of the material in the disk can also explain the small value of the strength of the torque noise observed from \f \citep{chakrabarty97}.

In the present work, following \citet{angelini95}, we have decomposed the strong emission-line complex centered on 1\,keV found in the \textit{Suzaku} X-ray spectrum of 4U\,1626-67 (see Fig.~\ref{lines}). In both \textit{Suzaku} observations the strongest line is the hydrogen-like Ne  Ly$\alpha$ line at $\sim$1.025\,keV, in agreement with the ASCA observations \citep{angelini95}.   The  broad feature at $\sim$1.48~keV  found in our observations is close to the expected energy of the H--like Ne recombination edge at 1.362\,keV. The measured 57\,eV broadening of the line at $\sim$1.48\,keV suggests that we may be also detecting the recombination continuum of Ne, as suggested by \cite{angelini95} for the ASCA observation.   We point out that the Mg H$\alpha$ transition occurs at energy  1.471\,keV (which has not been resolved by \textit{Suzaku}), therefore this Mg line may be blended with the recombination continuum of Ne. Furthermore, with \textit{Suzaku} an eighth  line centered at $\sim$1.34\,keV was detected, which we tentatively identify as the Mg He$\alpha$ (see Table~\ref{linesfit}).

During the spin--down period, comparing the strength  of the low-energy lines detected in  the Mar 2006 \textit{Suzaku} observation with previous measurements  by \textit{ASCA}, we see that the overall decrease is of the order of $\sim$2.5 (see Table~\ref{linesfit}). Before the
2008 torque reversal, X--ray missions  like  \textit{BeppoSAX} \citep{owens97},  \textit{Chandra} \citep{schulz01, krauss07}  and \textit{XMM} \citep{krauss07}  were able to resolve only up to 3 or 4 lines of this complex.  In general, the measurements of the intensity of those lines are in agreement with our results. During the new spin--up period after the 2008 torque reversal, the flux of the lines increased in general by a factor of $\sim$5 except the Ne~Ly$\alpha$ line at 1.021\,keV that increased by a factor of $\sim$8. This result is not surprising taking into account such a change in luminosity. Unfortunately, during a previous  spin-up period there was only one measurement of the $\sim$1\,keV complex by the \textit{Einstein} observatory  in 1979, and it was not resolved due to the lower data quality. The overall luminosity decrease associated with the 1990 torque reversal was a factor of $\sim$3   in the 0.5--10\,keV range \citep[and references therein]{angelini95}. \citet{krauss07} found a decrease of the equivalent widths of the   Ne\,He$\alpha$ 0.914\,keV and Ne\,Ly$\alpha$  1.021\,keV lines during the spin-down period. The behavior of the EW  for the O\,He$\alpha$ 0.568\,keV and O\,Ly$\alpha$ 0.653\,keV lines was more steady during the same period. With \textit{Suzaku} we found a large increase of the EW of the Ne\,Ly$\alpha$  1.021\,keV line after the 2008 torque reversal. The EW for this line reached almost the same value measured by ASCA in 1993. For the other lines the increase is less pronounced.

The strong emission-line complex at $\sim$1\,keV has been simply explained as being due to a substantial Ne overabundance \citep{angelini95}. Angelini et al. (1995) raised the possibility that 4U\,1626-67 is embedded in a shock heated gas from the supernova remmant that created the pulsar, but as they point out, the absence of strong lines from Si, S, and Fe is inconsistent with this idea. Instead, \cite{schulz01} proposed that this cool material originated in the accretion disk or the mass donor, so that its composition is then a clue to the nature of the companion. \cite{schulz01}  detected double-peaked emission lines and strong absorption edges of Ne and O in their \textit{Chandra} HETGS spectrum, thus pointing to the companion in 4U\,1626-67 being the chemically fractionated core of a C--O or O--Ne--Mg crystallized white dwarf. \cite{schulz01}  also  concluded that the strong edges are due to absorption in cool, metal-rich material local to the source.  \cite{krauss07} found that the double-peaked emission lines and the line ratios implied that they are formed in a high-density environment, suggesting  that they arise somewhere in the accretion disk. \cite{krauss07} also noted that their value for the temperature of the line formation region ($\geq10^6$ K) is characteristic of the highly ionized, optically thin outer layers of the accretion disk. We note that, in our work, single Gaussian line profiles  provided an acceptable fit to the observed line shapes, and  physical models like \textit{meka} or \textit{vmeka} did not yield  good fits.

\begin{acknowledgements}
  A.C.A. and M.H.F. acknowledge support from NASA  grants  NNX08AW06G and NNX11AE24G.
D.M.M. and K.P. acknowledge support from NASA grants NNX10AJ48G and NNX11AD41G. N.R.I. acknowledges support from the Program N21 of the Prezidium of RAS, Federal program "Scientific and pedagogical brain-power of
Russia" under the grant XXXVII--1.2.1 and NSH--1625.2012.2.
\end{acknowledgements}

\bibliographystyle{bibtex/aa}
\bibliography{elif}

\end{document}